# Valence-quark-core and meson-cloud effects on electromagnetic form factors of $\Delta$ baryons

Yutao Fan,[1] Shuai Gui,[1, *] Shixing Liu,[1, 2] Yongxin Guo,[1, 2] and Xuyang Liu[1, 2, †]

[1] *School of Physics, Liaoning University, Shenyang 110036, China*
[2] *Institute of Space Science and Technology, Liaoning University, Shenyang 110036, China*

Electromagnetic form factors and electromagnetic properties of $\Delta$ baryons are studied in the perturbative chiral quark model (PCQM) using quark wavefunctions predetermined from proton charge form factor data. Our predictions agree reasonably with lattice-QCD and other phenomenological results. For charged $\Delta^{++}$, $\Delta^{+}$, and $\Delta^{-}$, observables are dominated by the three-quark core, with small meson-cloud corrections superimposed. For neutral $\Delta^0$, the leading-order quark core contribution vanishes, so meson-cloud corrections fully determine its electromagnetic response. The PCQM gives a nonzero magnetic moment, a negative charge radius, and a sizable magnetic radius for $\Delta^0$. This work provides consistent, parameter-free predictions for $\Delta$ baryon structure and highlights the role of meson cloud fluctuations.

## I. INTRODUCTION

Electromagnetic form factors are key observables that provide direct access to the internal charge and magnetization distributions, and thus reveal the internal structure of baryons. Experimentally, however, their precise determination remains extremely challenging. High-precision measurements via electron scattering are strictly confined to stable nucleons [1, 2]. In contrast, the vast majority of baryon resonances possess extremely short lifetimes, rendering the direct experimental extraction of their electromagnetic form factors practically unfeasible at current facilities.

The $\Delta$ baryons $(\Delta^{++}, \Delta^{+}, \Delta^{0}, \Delta^{-})$, as the lightest members of the spin-3/2 decuplet, have long served as a benchmark for testing chiral dynamics and lattice QCD in the spin-3/2 sector. Experimental data exist only for the magnetic moments $\mu_B$ of $\Delta^{++}$ and $\Delta^{+}$, and even these are affected by considerable uncertainties [3]. With experimental avenues largely restricted, investigating the electromagnetic properties of these short-lived states relies fundamentally on theoretical simulations.

Lattice QCD (LQCD), as a first-principles approach, provides the most rigorous framework for baryon observables and serves as a theoretical benchmark where experimental data are unavailable. So far, LQCD has mainly provided results for the electromagnetic properties of the $\Delta$ baryons [4–9]. In Ref. [4], the electromagnetic form factors of the $\Delta^{+}$ were computed using quenched QCD, two flavors of dynamical Wilson quarks, and a mixed action with domain-wall valence quarks on staggered sea quarks, reaching pion masses down to approximately 350 MeV. A subsequent study extended this work and provided further results [5]. Refs. [6–9] calculated the magnetic moments of the $\Delta$ baryons. In particular, Ref. [9] reported a nonzero $\Delta^0$ magnetic moment of $-0.035(2)\,\mu_N$. Nevertheless, these simulations are performed at heavier-than-physical pion masses and have not been extrapolated to the physical point.

Besides LQCD, various phenomenological approaches have been employed to study the electromagnetic properties of the $\Delta$ baryon decuplet. The self-consistent SU(3) chiral quark-soliton model ($\chi$QSM) examined the electromagnetic form factors of the $\Delta$ resonance for $Q^2 \leq 1$ GeV$^2$, including $1/N_c$ rotational corrections. It also provides $\mu_{\Delta^{++}} = 3.65\,\mu_N$ and a nonzero $\mu_{\Delta^0} = -0.21\,\mu_N$, with a notably large magnetic radius for the $\Delta^0$ [10]. The covariant quark-diquark model (CQDM) calculated the electromagnetic form factors of the $\Delta$ baryons up to $Q^2 = 2$ GeV$^2$ [11], obtaining results in reasonable agreement with lattice QCD data for some observables. Assuming strict isospin symmetry, CQDM gives a vanishing $\Delta^0$ magnetic moment and identical radii for all charged $\Delta$ states. In contrast, chiral perturbation theory ($\chi$PT) [12, 13] suggests a nonzero $\mu_{\Delta^0}$. Other frameworks, including relativistic quark models (RQM) [14], chiral quark model ($\chi$QM) [15], chiral constituent-quark model ($\chi$CQM) [16], QCD sum rules [17], and $1/N_c$ expansion [18–20] also yield distinct predictions for magnetic moments, charge radii and magnetic radii. These theoretical predictions differ substantially, reflecting different meson-cloud treatments, and motivate the present calculation within an alternative meson-cloud scheme.

The perturbative chiral quark model (PCQM) is an effective tool to study the baryon structures and properties in the low-energy region [21–31], and provides a framework that explicitly incorporates both the valence-quark core and the meson cloud. In the PCQM, three relativistic valence quarks are confined by static scalar and vector potentials, while pseudoscalar mesons are introduced as dynamical degrees of freedom through a nonlinear $\sigma$ model Lagrangian, ensuring the preservation of chiral symmetry. In our previous works [32, 33], we numerically obtained the wavefunctions for a single valence quark consistent with the Cornell-like potential of the PCQM. Based on the predetermined quark wave-functions, the electroweak form factors and properties of baryon octets were estimated [32, 34–36]. Very recently, we further investigated the axial transition form factors for the semileptonic octet baryon decays $B' \to Bl\bar{\nu}$ [37] in the PCQM. These PCQM results are found to be in good agreement with experimental data and the lattice QCD values. Those studies also reveal that the meson cloud plays a significant role in baryon electroweak properties. In this work, we extend the PCQM to the $\Delta$ baryons and investigate the respective contributions of the valence-quark core and the meson cloud to the electromagnetic properties of these resonances.

* guishuailnu@163.com
† liuxuyang@lnu.edu.cn

The paper is organized as follows. In Sec. II, the theoretical framework of the PCQM is briefly introduced, and the analytical expressions for the electromagnetic form factors of $\Delta$ baryons are derived. Sec. III presents the numerical results and discussions. Finally, a summary and conclusions are given in Sec. IV.

## II. ELECTROMAGNETIC FORM FACTORS IN THE PERTURBATIVE CHIRAL QUARK MODEL

The underlying dynamics of the PCQM are dictated by the chiral effective Lagrangian. Following a unitary chiral transformation [**?** ], the PCQM Lagrangian can be derived in the Weinberg-type formulation,

$$\begin{aligned}\mathcal{L}_0(x) &= \bar{\psi}(x)\big[i\partial\!\!\!/ - \gamma^0 V(r) - S(r)\big]\psi(x) \\ &\quad -\frac{1}{2}\Phi_i(x)\big(\Box + M_\Phi^2\big)\Phi^i(x), \qquad (1)\\ \mathcal{L}_I^W(x) &= \frac{1}{2F}\partial_\mu\Phi_i(x)\bar{\psi}(x)\gamma^\mu\gamma^5\lambda^i\psi(x). \qquad (2)\end{aligned}$$

Here, $F = 88$ MeV denotes the pion decay constant at the chiral limit. $\Phi_i(x)$ are the octet meson fields. The quark field $\psi(x)$ is constructed as an SU(3) flavor triplet of the $u$, $d$, and $s$ quarks, and it could be expanded into

$$\psi(x) = \sum_\alpha b_\alpha u_\alpha(\vec{x})e^{-iE_\alpha t}, \qquad (3)$$

with $b_\alpha$ being the single quark annihilation creation operators. $u_\alpha$ are the set of quark wavefunctions in orbits $\alpha$. In general, the relativistic quark wavefunctions $u_\alpha(\vec{x})$ can be written as

$$u_\alpha(\vec{x}) = \begin{pmatrix} g_\alpha(r) \\ i\vec{\sigma}\cdot\hat{x}f_\alpha(r)\end{pmatrix}\chi_s\chi_f\chi_c, \qquad (4)$$

where $\chi_s$, $\chi_f$, and $\chi_c$ are the spin, flavor, and color wavefunctions of the quark, while $g_\alpha(r)$ and $f_\alpha(r)$ correspond to the upper and lower Dirac components, respectively. As established in our previous work [32], the ground-state quark wavefunction $u_0(\vec{x})$ was extracted by fitting the PCQM calculations of the proton charge form factor $G_E^p(Q^2)$ to the experimental measurements within the Sturmian basis expansion scheme.

In the PCQM, physical observables are calculated via the GellMann-Low theorem [38], wherein the charge form factor $G_E^B(Q^2)$ and the magnetic form factor $G_M^B(Q^2)$ for a baryon in the Breit frame are formulated as a perturbative expansion of time-ordered products over connected Feynman diagrams

$$\begin{aligned}&\chi^\dagger_{B_s}\chi_{B_s}G_E^B(Q^2) \\ &\quad = {}^B\langle\phi_0|\sum_{n=0}^{\infty}\frac{i^n}{n!}\int\delta(t)d^4xd^4x_1\ldots d^4x_n e^{-iq\cdot x} \\ &\qquad \times T\left[\mathcal{L}_I^W(x_1)\cdots\mathcal{L}_I^W(x_n)j^0(x)\right]|\phi_0\rangle^B, \qquad (5)\end{aligned}$$

$$\begin{aligned}&\chi^\dagger_{B_s}\frac{i\vec{\sigma}\times\vec{q}}{2m_B}\chi_{B_s}G_M^B(Q^2) \\ &\quad = {}^B\langle\phi_0|\sum_{n=0}^{\infty}\frac{i^n}{n!}\int\delta(t)d^4xd^4x_1\cdots d^4x_n e^{-iq\cdot x} \\ &\qquad \times T\left[\mathcal{L}_I^W(x_1)\cdots\mathcal{L}_I^W(x_n)\vec{j}(x)\right]|\phi_0\rangle^B, \qquad (6)\end{aligned}$$

where $\chi_{B_s}$ is the baryon spin wavefunction. $m_B$ is the physical baryon mass, and $\vec{\sigma}$ is the baryon spin matrix. The state vector $|\phi_0\rangle^B$ corresponds to the unperturbed three-quark states projected onto the respective baryon states. The EM current operator $j^\mu$ in Eqs. (5) and (6) reads

$$\begin{aligned}j^\mu &= \bar{\psi}\gamma^\mu\mathcal{Q}\psi + \bar{\psi}(\hat{Z}-1)\gamma^\mu\mathcal{Q}\psi \\ &\quad + \left[f_{3ij} + \frac{f_{8ij}}{\sqrt{3}}\right]\Phi_i\partial^\mu\Phi_j \\ &\quad + \left[f_{3ij} + \frac{f_{8ij}}{\sqrt{3}}\right]\frac{\Phi_j}{2F}\bar{\psi}\gamma^\mu\gamma^5\lambda_i\psi. \qquad (7)\end{aligned}$$

Here, $\mathcal{Q} = \mathrm{diag}(2/3, -1/3, -1/3)$ denotes the quark charge matrix, and $Z$ stands for the wavefunction renormalization constant. Following the renormalization scheme from Ref. [23], we derive the renormalization constant for non-strange quark $\hat{Z}$ as

$$\begin{aligned}\hat{Z} &= 1 - \frac{3}{4(2\pi F)^2}\int_0^\infty dk k^4 F_I^2(k^2) \\ &\quad \times\left[\frac{1}{\omega_\pi^3(k^2)} + \frac{2}{3\omega_K^3(k^2)} + \frac{1}{9\omega_\eta^3(k^2)}\right], \qquad (8)\end{aligned}$$

where $\omega_\Phi(k^2) = \sqrt{M_\Phi^2 + k^2}$ defines the energy dispersion relation of the intermediate Goldstone bosons, and the quark-meson vertex function $F_I(k)$ takes the form

$$\begin{aligned}F_I(k) &= 2\pi\int_0^\infty dr\int_0^\pi d\theta r^2\sin\theta e^{ikr\cos\theta} \\ &\quad \times\left[g(r)^2 + f(r)^2\cos 2\theta\right]. \qquad (9)\end{aligned}$$

Using the interaction Lagrangian $L_I^W(x)$ and the EM current $j^\mu$, we divide the EM form factors of $\Delta$ baryons up to the one-loop order into four Feynman diagrams, as shown in Fig. 1. The calculations are generalized to the full SU(3) flavor symmetry, incorporating meson-cloud contributions from $\pi$, kaon, and $\eta$ states. The corresponding SU(3) analytical expressions are therefore given below:

**(1) Three-quark core leading-order (LO) diagram**

$$G_E^B\left(Q^2\right)\big|_{LO}^{3q} = a_1^B \int_0^{2\pi} d\phi \int_0^\infty dr \int_0^\pi d\theta\, r^2 \sin\theta \times e^{iQr\cos\theta}\left[g(r)^2 + f(r)^2\right], \qquad (10)$$

$$G_M^B\left(Q^2\right)\big|_{LO}^{3q} = -b_1^B \frac{4\pi i m_B}{Q} \int_0^\infty dr \int_0^\pi d\theta r^2 \sin(2\theta) \times g(r) f(r) e^{iQr\cos\theta}. \qquad (11)$$

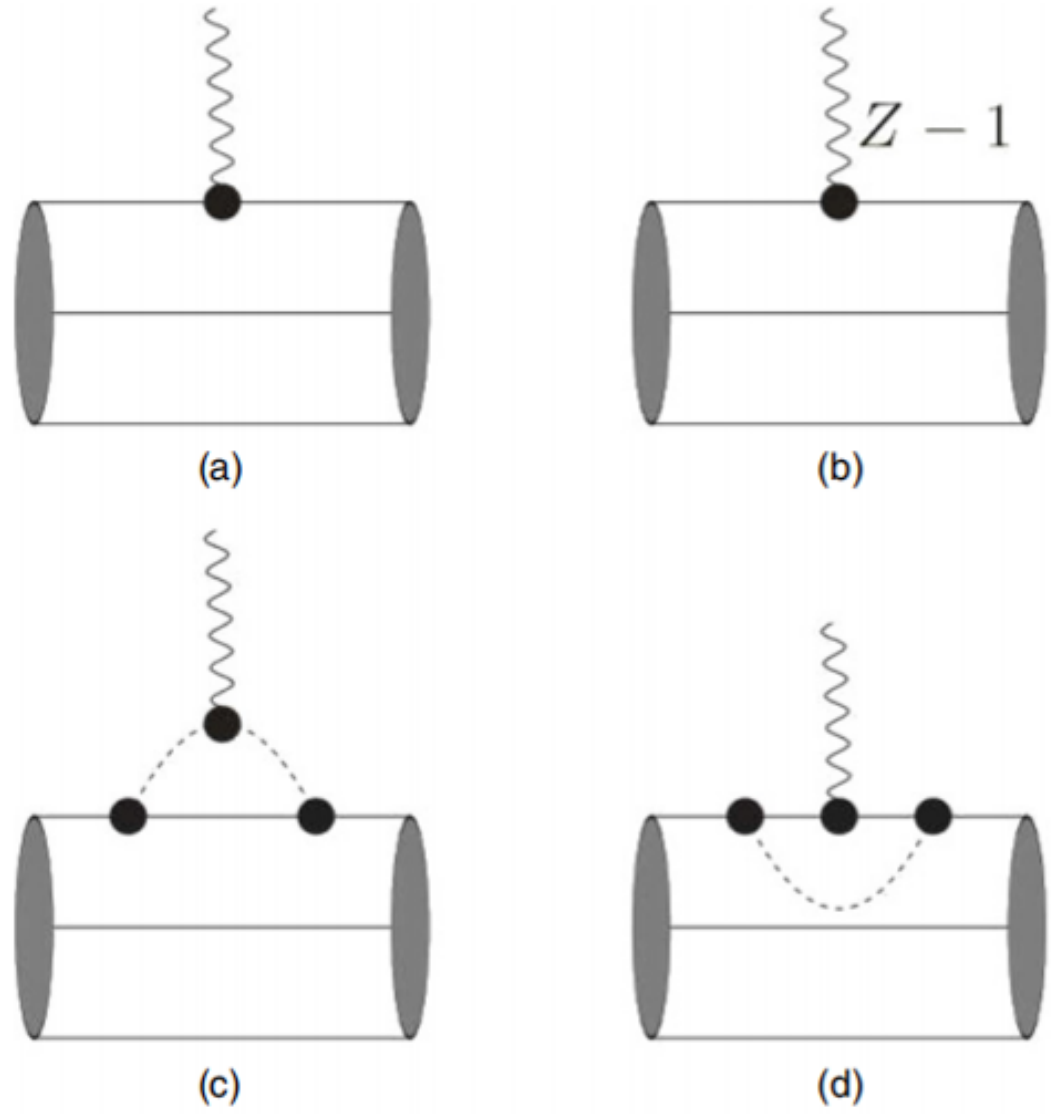

FIG. 1. Feynman diagrams contributing to the electromagnetic form factors of Δ baryons: (a) Three-quark diagram, (b) Three-quark counterterm diagram, (c) Meson cloud diagram, (d) Vertex correction diagram.

**(2) Three-quark core counterterm (CT) diagram**

$$G_E^B\left(Q^2\right)\big|_{CT}^{3q} = \left(\hat{Z}-1\right) G_E^B\left(Q^2\right)\big|_{LO}^{3q}, \qquad (12)$$

$$G_M^B\left(Q^2\right)\big|_{CT}^{3q} = \left(\hat{Z}-1\right) G_M^B\left(Q^2\right)\big|_{LO}^{3q}. \qquad (13)$$

**(3) Meson-cloud (MC) diagram**

$$G_E^B\left(Q^2\right)\big|_{MC} = \frac{1}{2(2\pi F)^2} \int_0^\infty dk \int_{-1}^1 dx\, k^2 \left(k^2 + kQx\right) F_I(k) F_I\left(k_+\right) \times \left[\frac{a_2^B}{\omega_\pi\left(k^2\right)\omega_\pi\left(k_+^2\right)\left[\omega_\pi\left(k^2\right)+\omega_\pi\left(k_+^2\right)\right]} + \frac{a_3^B}{\omega_K\left(k^2\right)\omega_K\left(k_+^2\right)\left[\omega_K\left(k^2\right)+\omega_K\left(k_+^2\right)\right]}\right], \qquad (14)$$

$$G_M^B\left(Q^2\right)\big|_{MC} = \frac{5m_B}{6(2\pi F)^2} \int_0^\infty dk\, k^4 \int_{-1}^1 dx \left(1-x^2\right) F_I(k) F_I\left(k_+\right) \left[\frac{b_2^B}{\omega_\pi^2\left(k^2\right)\omega_\pi^2\left(k_+^2\right)} + \frac{b_3^B}{\omega_K^2\left(k^2\right)\omega_K^2\left(k_+^2\right)}\right], \qquad (15)$$

where

$$k_+ = \sqrt{k^2 + Q^2 + 2k\sqrt{Q^2}x}. \qquad (16)$$

**(4) Vertex-correction (VC) diagram**

$$G_E^B\left(Q^2\right)\big|_{VC} = \frac{1}{2(2\pi F)^2} \int_0^\infty dk\, k^4 F_I^2(k)\, G_E^B\left(Q^2\right)\big|_{LO}^{3q} \times \left[\frac{a_4^B}{\omega_\pi^3(k^2)} + \frac{a_5^B}{\omega_K^3(k^2)} + \frac{a_6^B}{\omega_\eta^3(k^2)}\right], \qquad (17)$$

$$G_M^B\left(Q^2\right)\big|_{VC} = \frac{1}{2(2\pi F)^2} \int_0^\infty dk\, k^4 F_I^2(k)\, G_M^B\left(Q^2\right)\big|_{LO}^{3q} \times \left[\frac{b_4^B}{\omega_\pi^3(k^2)} + \frac{b_5^B}{\omega_K^3(k^2)} + \frac{b_6^B}{\omega_\eta^3(k^2)}\right]. \qquad (18)$$

In the above equations, $a_i^B$ and $b_i^B$ represent the matrix elements of the flavor and spin operators acting on the baryon states, which are listed in Tables I and II respectively. These quantities could be computed by means of the PCQM projection formalism

$$\chi_{f'}^\dagger \chi_{s'}^\dagger I^{f'f} J^{s's} \chi_s^\dagger \chi_f^\dagger \xrightarrow{Proj.} \langle B| \sum_{i=1}^{3} (IJ)^{(i)} |B\rangle, \qquad (19)$$

where $I$ and $J$ are operators which act in flavor and spin spaces, respectively, and the baryon states $|B\rangle$ are constructed in the framework of the $SU(6)$ spin-flavor and $SU(3)$ color symmetry.

TABLE I. The constants $a_i^B$ for the charge form factors $G_E^B(Q^2)$.

| Baryon | $a_1^B$ | $a_2^B$ | $a_3^B$ | $a_4^B$ | $a_5^B$ | $a_6^B$ |
|---|---|---|---|---|---|---|
| $\Delta^{++}$ | 2 | 3 | 3 | 0 | -1/2 | 1/6 |
| $\Delta^{+}$ | 1 | 1 | 2 | 1/2 | -1 | 1/6 |
| $\Delta^{0}$ | 0 | -1 | 1 | 0 | 0 | 0 |
| $\Delta^{-}$ | -1 | -3 | 0 | -3/2 | 1 | 1/6 |

TABLE II. The constants $b_i^B$ for the magnetic form factors $G_M^B(Q^2)$.

| Baryon | $b_1^B$ | $b_2^B$ | $b_3^B$ | $b_4^B$ | $b_5^B$ | $b_6^B$ |
|---|---|---|---|---|---|---|
| $\Delta^{++}$ | 2 | 9/5 | 9/5 | 0 | 1/2 | -1/6 |
| $\Delta^{+}$ | 1 | 3/5 | 6/5 | -1/2 | 1 | -1/6 |
| $\Delta^{0}$ | 0 | -3/5 | 3/5 | 0 | 0 | 0 |
| $\Delta^{-}$ | -1 | -9/5 | 0 | 3/2 | -1 | -1/6 |

## III. NUMERICAL RESULTS AND DISCUSSION

In this section, we evaluate our numerical results for the electromagnetic form factors and static properties of the $\Delta$ baryons. As discussed in Sec. II, the ground-state quark wavefunction employed in all calculations was predetermined by fitting the PCQM prediction of the $G_E^p(Q^2)$ to the experimental data, and it is adopted here without further adjustment. Consequently, all results below are parameter-free.

In Fig. 2, we first present the $Q^2$-dependence of the $\Delta^+$ electromagnetic form factors $G_E^{\Delta^+}(Q^2)$ and $G_M^{\Delta^+}(Q^2)$ for $0 \leqslant Q^2 \leqslant 2.5$ GeV$^2$, in comparison with the recent lattice QCD data [4] at the pion masses $m_\pi = 353$, 384, and 410 MeV. Overall, the total PCQM curves for both $G_E^{\Delta^+}(Q^2)$ and $G_M^{\Delta^+}(Q^2)$ are qualitatively consistent with the lattice trend but lie slightly below the lattice points. This discrepancy arises because the lattice QCD simulations of Ref. [4] are performed at heavier-than-physical pion masses, whereas the present PCQM calculation uses the physical pion mass. The lattice data in Fig. 2 indicate a trend that the electromagnetic form factors of $\Delta^+$ decrease with decreasing pion mass. This suggests that the lattice-QCD form factor magnitudes should be reduced upon chiral extrapolation to the physical pion mass, supporting our PCQM predictions.

Fig. 2 also decomposes the total PCQM results into three-quark core leading order (LO), counterterm (CT), meson cloud (MC), and vertex correction (VC) contributions. The LO dominates both charge and magnetic form factors across the entire $Q^2$ range, while the chiral corrections are rather small. The MC diagram gives a minor positive contribution at low $Q^2$, which is nearly canceled by the negative CT term, but the VC term is negligible. This decomposition indicates that the electromagnetic structure of $\Delta^+$ is essentially governed by the valence quark distribution, with the meson cloud playing only a secondary role.

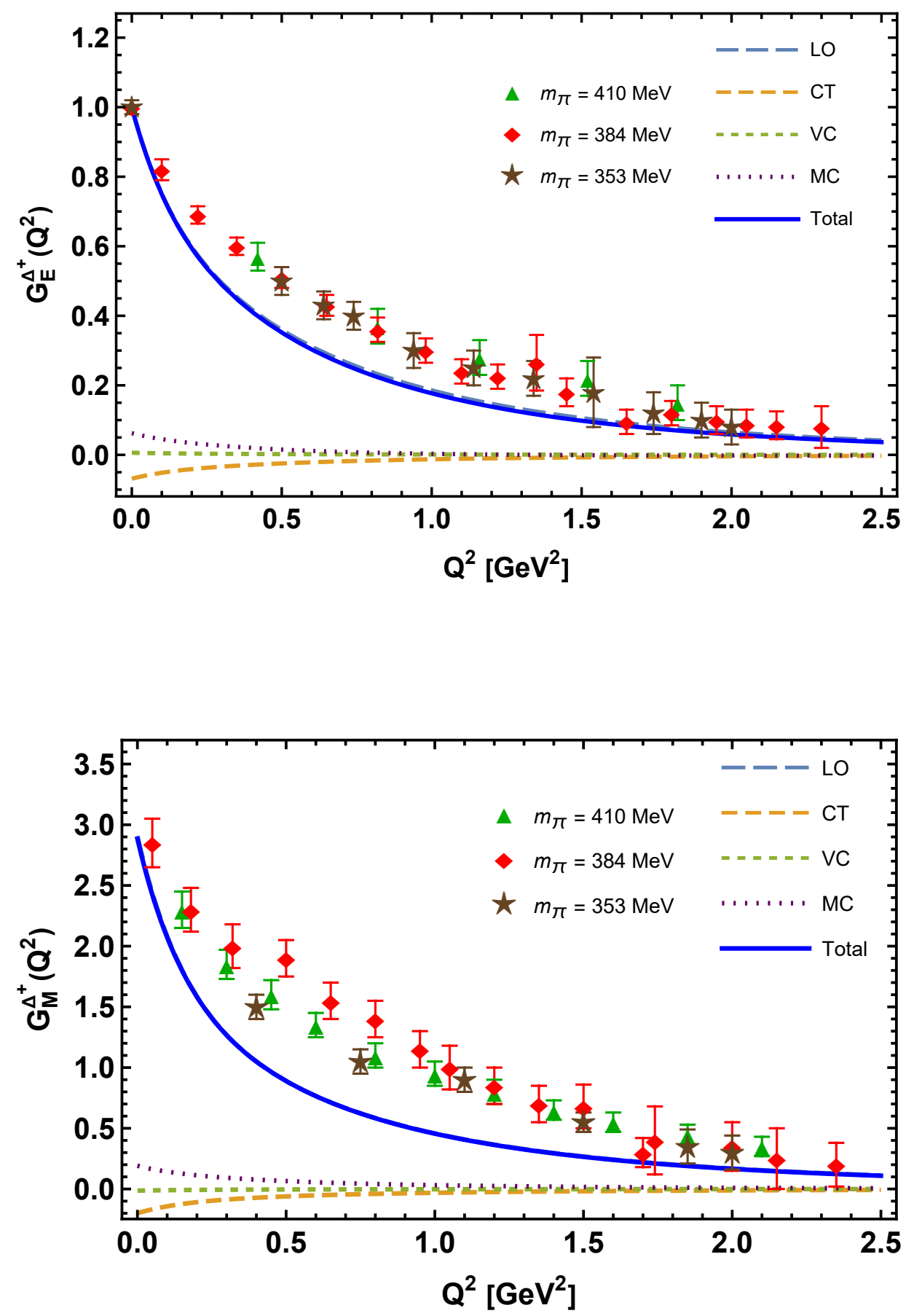

FIG. 2. The PCQM results of the $\Delta^+$ baryon electromagnetic form factors comparing with the lattice QCD data [4].

Next, the PCQM predictions for the electromagnetic form factors of the $\Delta^{++}$, $\Delta^-$ and $\Delta^0$ baryons are compiled in Fig. 3. Similar to the $\Delta^+$ results, the form factors for the charged $\Delta^{++}$ and $\Delta^-$ show a smooth dipole-like falloff with $Q^2$. The LO overwhelmingly dominates the total result across the entire $Q^2$ range, while the meson clouds play only a minor role. In contrast to the charged states, the neutral $\Delta^0$ baryon exhibits distinctly different behavior in both its charge and magnetic form factors. As shown in Fig. 3, $G_E^{\Delta^0}(Q^2)$ and $G_M^{\Delta^0}(Q^2)$ are substantially smaller in magnitude than those of the charged states. Since the three-quark core of the $\Delta^0$ carries no net charge, one has $a_1^{\Delta^0} = 0$ as listed in Table I, and consequently the LO charge contribution vanishes identically for all $Q^2$. $G_E^{\Delta^0}(Q^2)$ is governed not by the valence quark core but rather by the meson-cloud fluctuations. The decomposition reveals that the VC diagram, which played only a minor role for the charged cases, now becomes the dominant source of the charge response, while the MC term provides the main cancellation. For $G_M^{\Delta^0}(Q^2)$, the LO contribution also vanishes in the PCQM owing to the fully symmetric spin-flavor structure of the decuplet wavefunction, corresponding to $b_1^{\Delta^0} = 0$ in Table II. Thus, $G_M^{\Delta^0}(Q^2)$ is also governed by meson-cloud fluctuations. These results, obtained without any parameter adjustment, demonstrate that the PCQM consistently describes the influence of chiral meson-cloud fluctuations in $\Delta$ baryons.

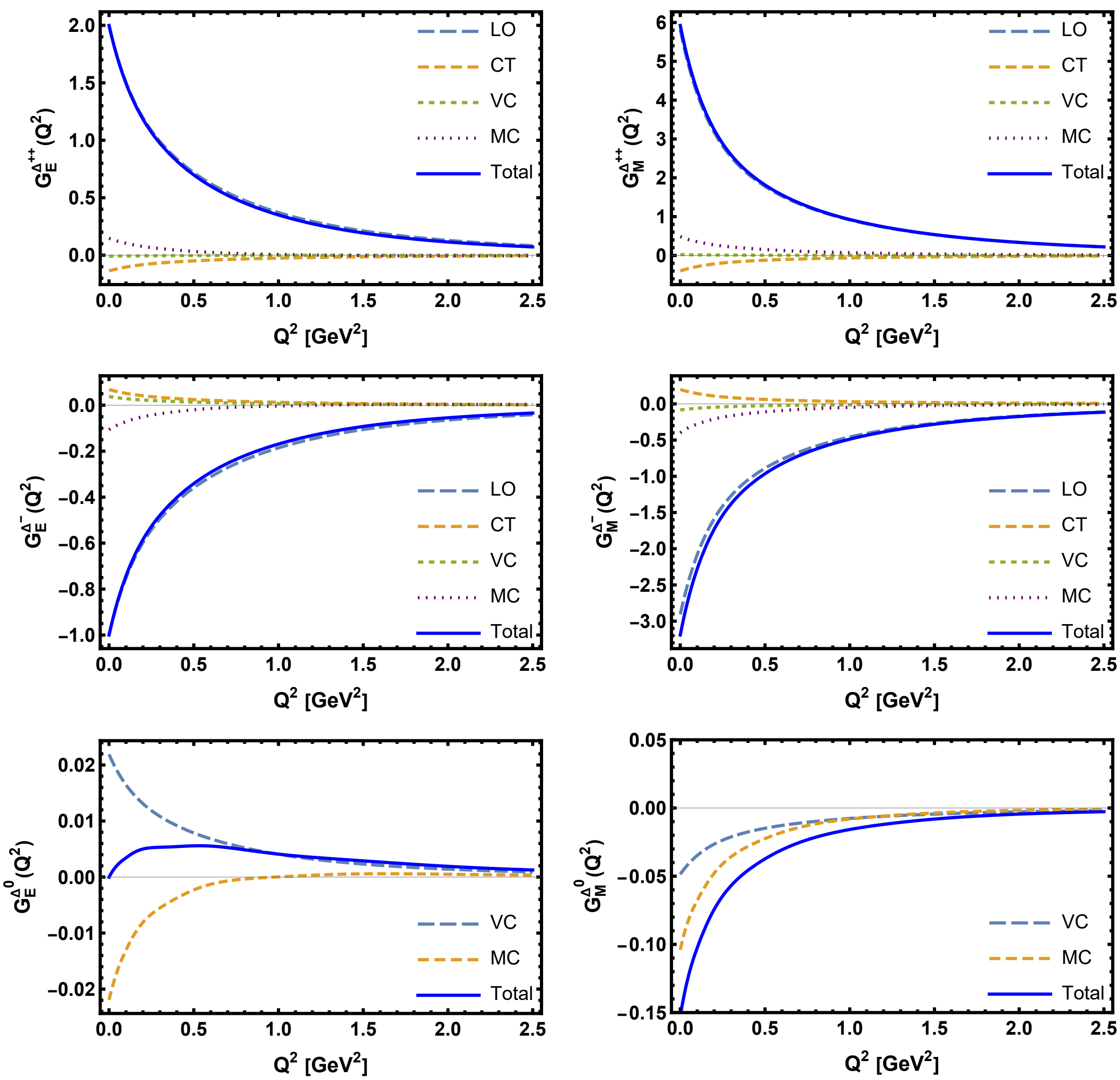

FIG. 3. The PCQM results for the charge and magnetic form factors of the $\Delta^{++}$, $\Delta^-$, and $\Delta^0$.

Finally, we turn to the static electromagnetic properties of the $\Delta$ baryons, which are intimately related to the behaviors of electromagnetic form factors at $Q^2 = 0$. These observables could complement the full $Q^2$-dependence of the form factors discussed above. In Table III, the magnetic moments $\mu_B = G_M^B(0)$ of the $\Delta$ baryons are summarized in units of $\mu_N$ together with the available experimental data [3], lattice QCD extractions [4–9] and results from various theoretical models [10–12, 14, 15, 17, 18] for comparison. The PCQM predictions at the physical pion mass are in good agreement with the experimental data or the lattice QCD estimates. In particular, we obtain $\mu_{\Delta^{++}} = 5.923$ and $\mu_{\Delta^+} = 2.885$, both of which lie well within the experimental error bands, and are also close to the $\chi$PT [12] and the $1/N_c$ expansion [18] predictions. For the $\Delta^0$, the PCQM predicts a small but nonzero magnetic moment $\mu_{\Delta^0} = -0.153$. Although the $\Delta^0$ shares the same *udd* valence-quark content as neutron, its magnetic moment is much smaller in magnitude. This difference arises from the fully symmetric spin-flavor wavefunction of the decuplet. As discussed above, this structure leads to an exact cancellation of the LO contribution, so the magnetic response is entirely generated by meson cloud fluctuations. The PCQM result for $\mu_{\Delta^0}$ is consistent in magnitude with the lattice QCD value [9] and $\chi$PT estimate [12], confirming that the $\Delta^0$ magnetic moment is indeed determined by the meson-cloud dynamics.

TABLE III. The PCQM numerical results for $\Delta$ baryon magnetic moments in units of $\mu_N$, compared with PDG experimental data[3] and various theoretical predictions from LQCD [4–9],$\chi$QSM [10], CQDM [11], $\chi$PT [12], RQM [14], $\chi$QM [15], QCD sum rules [17], and $1/N_c$ expansion [18].

| $\mu/\mu_N$ | $\Delta^{++}$ | $\Delta^{+}$ | $\Delta^{0}$ | $\Delta^{-}$ |
|---|---|---|---|---|
| PCQM | 5.923 | 2.885 | −0.153 | −3.191 |
| PDG [3] | $6.14 \pm 0.51$ | $2.7^{+1.0}_{-1.3} \pm 1.5 \pm 3$ | ··· | ··· |
| LQCD [4, 5] | ··· | 1.91(16) | ··· | ··· |
| LQCD [6] | 3.20(56) | 1.60(28) | 0 | −1.60(28) |
| LQCD [7] | 3.70(12) | 2.40(6) | ··· | −1.85(6) |
| LQCD [8] | 4.91(61) | 2.46(31) | 0 | −2.46(31) |
| LQCD [9] | 5.24(18) | 0.97(8) | −0.035(2) | −2.98(19) |
| $\chi$QSM [10] | 3.65 | 1.72 | −0.21 | −2.14 |
| CQDM [11] | 4.80 | 2.40 | 0 | −2.40 |
| $\chi$PT [12] | 6.04(13) | 2.84(2) | −0.36(9) | −3.56(20) |
| RQM [14] | 4.76 | 2.38 | 0 | −2.38 |
| $\chi$QM [15] | 6.93 | 3.47 | 0 | −3.47 |
| QCD sum rules [17] | 4.13(1.30) | 2.07(65) | 0 | −2.07(65) |
| $1/N_c$ expansion [18] | 5.9(4) | 2.9(2) | ··· | −2.9(2) |

It is worth noting that the PCQM results for $\mu_B$ in Table III also show an exact linear decrease across the $\Delta$ multiplet, with a constant difference $-3.038\ \mu_N$ between adjacent states. This linearity reflects the SU(2) isospin symmetry preserved in the model. Among the theoretical results compiled in Table III, several quark model calculations such as CQDM, RQM, $\chi$QM, QCD sum rules and certain lattice-QCD data also satisfy the isospin-linear relation but produce $\mu_{\Delta^0} = 0$, as the meson cloud degrees of freedom are absent or suppressed. Only $\chi$PT and $\chi$QSM provide similar conclusions. These observations confirm that the valence-quark core determines the isospin-linear pattern, while the nonzero $\mu_{\Delta^0}$ originates from meson-cloud fluctuations.

Table IV and Table V give the charge and magnetic radii, respectively, which are derived from the slopes of the electromagnetic form factor at $Q^2 = 0$. For the charged $\Delta^{++}$, $\Delta^{+}$, and $\Delta^{-}$ states, the predicted charge and magnetic radii exhibit comparable magnitudes. This arises since their electromagnetic structure is dominated by the three-quark core contribution. Owing to isospin symmetry, the radial wave functions of $u$ and $d$ quarks are identical, leading to similar spatial extensions of the valence-quark cores. Meson-cloud corrections produce only modest shifts and do not substantially alter the overall spatial size. The charge radii for $\Delta^{++}$ and $\Delta^{-}$ are 0.761 fm$^2$ and 0.782 fm$^2$, respectively, in reasonable agreement with $\chi$QM [15] and $1/N_c$ expansion [20] predictions, while somewhat larger than the $\chi$PT and $\chi$CQM estimates. The corresponding magnetic radii fall within a similar numerical range and are qualitatively consistent with other theoretical predictions.

In sharp contrast, for the neutral $\Delta^0$, our PCQM yields a small negative charge radius $\langle r_E^2 \rangle^{\Delta^0} = -0.014$ fm$^2$, compatible with predictions from $\chi$PT [13] and the $\chi$QSM [10], whereas the lattice QCD, CQDM, and $\chi$QM give zero. A nonzero magnetic radius for $\Delta^0$ is also predicted within the PCQM. As already discussed from the full-$Q^2$ form factor decomposition for $\Delta^0$ shown in Fig. 3, the leading-order core contribution vanishes, and meson-cloud fluctuations therefore entirely determine its static electromagnetic observables.

TABLE IV. The PCQM numerical results for charge radii of $\Delta$ baryons in units of $\mathrm{fm}^2$, compared with those from LQCD [4, 5], $\chi$QSM [10], CQDM [11], $\chi$PT [13], $\chi$QM [15], $\chi$CQM [16], and $1/N_c$ expansion [19, 20].

| $\langle r_E^2 \rangle/\mathrm{fm}^2$ | $\Delta^{++}$ | $\Delta^{+}$ | $\Delta^{0}$ | $\Delta^{-}$ |
|---|---|---|---|---|
| PCQM | 0.761 | 0.754 | −0.014 | 0.782 |
| LQCD [4, 5] | ⋯ | 0.641(22) | ⋯ | ⋯ |
| $\chi$QSM [10] | 0.826 | 0.792 | −0.069 | 0.930 |
| CQDM [11] | 0.894 | 0.894 | 0 | 0.894 |
| $\chi$PT [13] | 0.30(11) | 0.29(10) | −0.02(1) | 0.33(11) |
| $\chi$QM [15] | 0.77 | 0.77 | 0 | 0.77 |
| $\chi$CQM [16] | 0.43 | 0.43 | 0 | 0.43 |
| $1/N_c$ [19] | 1.048 | 1.101 | 0.105 | 0.891 |
| $1/N_c$ [20] | 0.783 | 0.783 | 0 | 0.783 |

TABLE V. The PCQM numerical results for magnetic radii of $\Delta$ baryons in units of $\mathrm{fm}^2$, compared with those from $\chi$QSM [10], CQDM [11], $\chi$PT [13], and $\chi$QM [15].

| $\langle r_M^2 \rangle/\mathrm{fm}^2$ | $\Delta^{++}$ | $\Delta^{+}$ | $\Delta^{0}$ | $\Delta^{-}$ |
|---|---|---|---|---|
| PCQM | 0.864 | 0.858 | 1.163 | 0.887 |
| $\chi$QSM [10] | 0.587 | 0.513 | 1.786 | 0.764 |
| CQDM [11] | 0.827 | 0.827 | 0 | 0.827 |
| $\chi$PT [13] | 0.61(15) | 0.64(14) | 0.07(12) | 0.55(19) |
| $\chi$QM [15] | 0.62 | 0.62 | 0 | 0.62 |

## IV. SUMMARY AND CONCLUSIONS

In this work, we systematically investigate the electromagnetic properties of $\Delta$ baryons in the PCQM. Using the quark wavefunctions predetermined by fitting the proton charge form factor data, all calculations for $\Delta$ baryons are carried out without additional free parameters. We compute both the $Q^2$-dependent electromagnetic form factors and static electromagnetic observables, including magnetic moments as well as charge and magnetic radii. Our predictions show reasonable overall agreement with the available lattice QCD simulations and results from various phenomenological approaches.

Our results show that the electromagnetic structure of charged $\Delta^{++}$, $\Delta^{+}$, and $\Delta^{-}$ baryons is dominated by the leading-order three-quark core contribution. Meson cloud corrections are small and merely provide minor modifications to the electromagnetic observables. In contrast, the leading-order quark core contribution vanishes for the neutral $\Delta^0$. Even though meson cloud corrections remain small for all $\Delta$ states across the whole $Q^2$ range, these small corrections become the sole source of $\Delta^0$ electromagnetic response. Consequently, the PCQM predicts a nonzero magnetic moment, a negative charge radius, and a finite magnetic radius for $\Delta^0$.

This work reveals the physical role of meson-cloud corrections in $\Delta$ baryon structure, and the parameter free predictions provide useful benchmarks for future theoretical studies.

## ACKNOWLEDGMENTS

This work is supported by the National Natural Science Foundation of China (Grants 12232009 and 12372002), the 2024 Fundamental Research Program of Liaoning University (Scientific Research Platform Construction Project LNU202404), and the Basic Scientific Research Project for Universities of Liaoning Provincial Department of Education (2026)–Innovation Team Project for High-End Talents (Grant No. LJ222610140003).

[1] C. F. Perdrisat, V. Punjabi, and M. Vanderhaeghen, Prog. Part. Nucl. Phys. **59**, 694 (2007).
[2] V. Punjabi, C. F. Perdrisat, M. K. Jones, E. J. Brash, and C. E. Carlson, Eur. Phys. J. A **51**, 79 (2015).
[3] F. Takahashi *et al.* (Particle Data Group), Int. J. Mod. Phys. A **2026**, 2630011 (2026).
[4] C. Alexandrou, T. Korzec, G. Koutsou, T. Leontiou, C. Lorcé, J. W. Negele, V. Pascalutsa, A. Tsapalis, and M. Vanderhaeghen, Phys. Rev. D **79**, 014507 (2009).
[5] C. Alexandrou, T. Korzec, G. Koutsou, J. W. Negele, and Y. Proestos, Phys. Rev. D **82**, 034504 (2010).
[6] S. Boinepalli, D. B. Leinweber, P. J. Moran, A. G. Williams, J. M. Zanotti, and J. B. Zhang, Phys. Rev. D **80**, 054505 (2009).
[7] C. Aubin, K. Orginos, V. Pascalutsa, and M. Vanderhaeghen, Phys. Rev. D **79**, 051502 (2009).
[8] D. B. Leinweber, T. Draper, and R. M. Woloshyn, Phys. Rev. D **46**, 3067 (1992).
[9] F. X. Lee, R. Kelly, L. Zhou, and W. Wilcox, Phys. Lett. B **627**, 71 (2005).
[10] J.-Y. Kim and H.-C. Kim, Eur. Phys. J. C **79**, 570 (2019).
[11] J. Q. Wang, D. Fu, and Y. Dong, Eur. Phys. J. C **84**, 79 (2024).
[12] L. S. Geng, J. Martin Camalich, and M. J. Vicente Vacas, Phys. Rev. D **80**, 034027 (2009).
[13] H.-S. Li, Z.-W. Liu, X.-L. Chen, W.-Z. Deng, and S.-L. Zhu, Phys. Rev. D **95**, 076001 (2017).
[14] F. Schlumpf, Phys. Rev. D **48**, 4478 (1993).

[15] G. Wagner, A. J. Buchmann, and A. Faessler, J. Phys. G: Nucl. Part. Phys. **26**, 267 (2000).
[16] K. Berger, R. F. Wagenbrunn, and W. Plessas, Phys. Rev. D **70**, 094027 (2004).
[17] F. X. Lee, Phys. Rev. D **57**, 1801 (1998).
[18] M. A. Luty, J. March-Russell, and M. J. White, Phys. Rev. D **51**, 2332 (1995).
[19] R. Flores-Mendieta and M. A. Rivera-Ruiz, Phys. Rev. D **92**, 094026 (2015).
[20] A. J. Buchmann and R. F. Lebed, Phys. Rev. D **67**, 016002 (2003).
[21] V. E. Lyubovitskij, T. Gutsche, A. Faessler, and E. G. Drukarev, Phys. Rev. D **63**, 054026 (2001).
[22] V. E. Lyubovitskij, T. Gutsche, A. Faessler, and R. Vinh Mau, Phys. Lett. B **520**, 204 (2001).
[23] V. E. Lyubovitskij, T. Gutsche, A. Faessler, and R. Vinh Mau, Phys. Rev. C **65**, 025202 (2002).
[24] V. E. Lyubovitskij, P. Wang, T. Gutsche, and A. Faessler, Phys. Rev. C **66**, 055204 (2002).
[25] K. Pumsa-ard, V. E. Lyubovitskij, T. Gutsche, A. Faessler, and S. Cheedket, Phys. Rev. C **68**, 015205 (2003).
[26] T. Inoue, V. E. Lyubovitskij, T. Gutsche, and A. Faessler, Phys. Rev. C **69**, 035207 (2004).
[27] S. Cheedket, V. E. Lyubovitskij, T. Gutsche, A. Faessler, K. Pumsa-ard, and Y. Yan, Eur. Phys. J. A **20**, 317 (2004).
[28] K. Khosonthongkee, V. E. Lyubovitskij, T. Gutsche, A. Faessler, K. Pumsa-ard, S. Cheedket, and Y. Yan, J. Phys. G: Nucl. Part. Phys. **30**, 793 (2004).
[29] Y. Dong, A. Faessler, T. Gutsche, J. Kuckei, V. E. Lyubovitskij, K. Pumsa-ard, and P. Shen, J. Phys. G: Nucl. Part. Phys. **32**, 203 (2006).
[30] C. Dib, A. Faessler, T. Gutsche, S. Kovalenko, J. Kuckei, V. E. Lyubovitskij, and K. Pumsa-ard, J. Phys. G: Nucl. Part. Phys. **32**, 547 (2006).
[31] A. Faessler, T. Gutsche, V. E. Lyubovitskij, and C. Oonariya, J. Phys. G: Nucl. Part. Phys. **35**, 025005 (2008).
[32] X. Y. Liu, K. Khosonthongkee, A. Limphirat, and Y. Yan, J. Phys. G: Nucl. Part. Phys. **41**, 055008 (2014).
[33] X. Y. Liu, Z. J. Liu, A. Limphirat, K. Khosonthongkee, and Y. Yan, Ann. Phys. **388**, 114 (2018).
[34] X. Y. Liu, K. Khosonthongkee, A. Limphirat, P. Suebka, and Y. Yan, Phys. Rev. D **91**, 034022 (2015).
[35] X. Y. Liu, D. Samart, K. Khosonthongkee, A. Limphirat, K. Xu, and Y. Yan, Phys. Rev. C **97**, 055206 (2018).
[36] X. Y. Liu, A. Limphirat, K. Xu, D. Samart, K. Khosonthongkee, and Y. Yan, Eur. Phys. J. A **55**, 218 (2019).
[37] X. Y. Liu, A. Limphirat, K. Xu, Z. Zhao, K. Khosonthongkee, and Y. Yan, Phys. Rev. D **107**, 074006 (2023).
[38] M. Gell-Mann and F. Low, Phys. Rev. **84**, 350 (1951).